\documentstyle[prd,aps,preprint,psfig]{revtex}

\def\pr{\,{Phys. Rev. }}

\begin{document}
\newcommand{\kpara}{k_{\parallel}}
\newcommand{\kperp}{k_{\perp}}
\newcommand{\me}{m_e}
\newcommand{\order}[1]{{\cal O}(#1)}
\newcommand{\vect}[1]{\mbox{\boldmath $#1$}}

\tighten
\draft

\preprint{
\noindent
\begin{minipage}[t]{3in}
\begin{flushright}
YITP-01-11\\
astro-ph/0102225 \\
February 2001
 \\
\end{flushright}
\end{minipage}
}

\title{Polarization tensors in a strong magnetic field}
\author{Kazunori Kohri}
\address{Yukawa Institute for Theoretical Physics, Kyoto University,
Kyoto, 606-8502, Japan}
\author{Shoichi Yamada}
\address{Institute of Laser Engineering (ILE), 
Osaka University, Osaka 565-0871,
Japan}
\date{\today}

\maketitle

\begin{abstract}
The vacuum polarization tensor in strong external magnetic
fields has been evaluated numerically for various strengths of
magnetic fields and momenta of photons under the threshold of 
the $e^{\pm}$ pair creation. The fitting formula has been obtained 
which reproduces the calculated results within 10 \% of error. 
The proper time method is employed further to obtain the retarded 
polarization tensor for finite temperature plasmas.
\end{abstract}



\section{Introduction}
\label{sec:Intro}

The strong magnetic field is attracting attentions of astrophysicists
these days. It has been known that the magnetized vacuum shows 
interesting features as the magnetic field strength exceeds a critical 
value $B_{c} = m_{e}^{2}/e \sim 4\times 10^{13}$G~\cite{du,lai}. Since 
this value is so large even in the universe compared, for example, with
the canonical magnetic field of $10^{12}$G for a pulsar, that it was
supposed that this was a subject of academic interest only. This has 
been changing drastically recently.

Some observations~\cite{ko} suggest the existence of neutron stars with 
a magnetic field far greater ($\sim 10^{15}$G) than the canonical one 
for the observed pulsars, and they are called a magnetar as dubbed by 
Duncun and Thompson~\cite{dt}. As the reality of very large magnetic 
fields looms, some researchers speculated further that some other 
peculiar extraterrestrial phenomena might also involve a very strong
magnetic fields. Among them are gamma ray bursts and
hypernovae~\cite{pa}. It is supposed for the latter that a jet is 
somehow produced in these objects and the dipolar magnetic
field is playing an essential role for that. On the other hand, some 
models for the gamma ray bursts are employing the magnetic fields to 
extract enormous energy of the phenomenon itself~\cite{bz}. 
Furthermore, the evidence of generic asymmetry for collapse-driven 
supernovae has been accumulated~\cite{wh}, and the strong magnetic 
field might have some implications for the ordinary supernova if the 
magnetar as observed is the end product of the supernova explosion 
and the observed asymmetry is a dynamical consequence involving the 
strong magnetic field~\cite{sy}. The fast proper motion of young
pulsars might be explained by the combination of the strong magnetic 
fields and some processes such as neutrino oscillations, 
for example~\cite{ks}. The explosion mechanism of the supernova will 
be changed substantially as well as nucleosynthesis therein.

It is, therefore, not only of academic interest to consider the features of
the strongly magnetized vacuum. In particular, the quantum electrodynamical
processes are most important, since those objects quoted above are 
mostly observed by electromagnetic waves, though the weak interactions 
are no less important~\cite{ks,hl}. The study of the strongly magnetized
vacuum has a long history, though. Adler~\cite{ad}, for example, gave detailed
formulations for the polarization tensor and the photon splitting rate 
as well as some useful analytic expressions for limit cases (see
also~\cite{lai,melrose_stoneham,mz} and the references therein for many other 
contributions). Astrophysicists have used these approximate formulae 
for their model building~\cite{td,bm}.

Those analytic expressions are approximate ones, though, valid for some
limit cases such as the strong or weak magnetic field limits and the 
zero photon energy limit. It appears that we are lacking the
complementary numerical evaluations of the polarization tensor for 
the intermediate values of magnetic field strength and/or photon
energy. It is the purpose of the paper to fill this gap
and give the interpolation formula based on the fitting to the numerical
integrations. Our interest is, however, also directed to the dispersive
relation of photon in the strongly magnetized plasma. In the last 
section we will extend the Schwinger's formulation and give the 
expression for the retarded polarization tensor for finite 
temperature plasmas.

\section{Vacuum polarization in a strong magnetic field}
\label{sec:Vacuum_polarization}

\begin{figure}
   \begin{center}
     \centerline{\psfig{figure=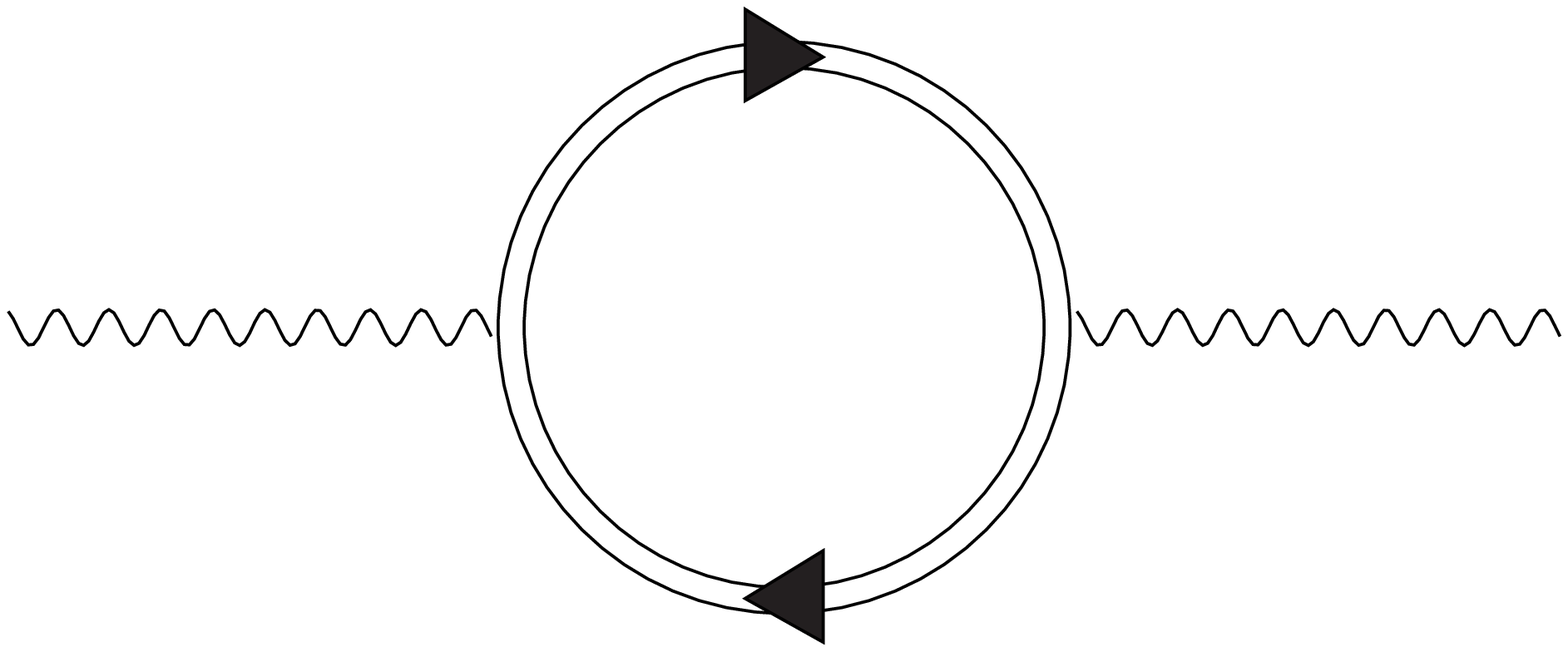,width=13cm}}
       \caption{
       Vacuum polarization in a strong magnetic field. The double
       solid line denotes the electron propagator which includes all
       contributions of the vertices from the  magnetic field.
       }
       \label{fig:vac_mag}
   \end{center}
\end{figure}

Using Schwinger's proper-time method~\cite{schwinger}, we obtain the
vacuum polarization tensor (see Fig.~\ref{fig:vac_mag}) in a strong
magnetic field~\cite{melrose_stoneham} expanded as,
\begin{equation}
    \label{eq:polarization_tensor}
    \Pi^{\mu\nu}(k, B) = \sum_{i = 0}^2 G_i f_i^{\mu\nu},
\end{equation}
where $k = (\omega, \kperp, 0, \kpara)$ is the energy-momentum
4-vector of photon, $\omega$ the photon energy, $\kperp =
\omega\sin\theta$, $\kpara = \omega\cos\theta$, $theta$ the direction 
of 3-momentum with respect to the magnetic field, and $B$ is the
magnetic field. $G_i$ is given by the following equations,
\begin{equation}
    \label{eq:gi}
    G_i = \frac{e^2}{(2\pi)^2}L\int^{\infty}_0 \!\! d\alpha \int^1_0 
    \! d\beta \  
    \gamma_i \, E(\alpha, \beta, k, L),
\end{equation}
with
\begin{eqnarray}
    \label{eq:gamma_i}
    \gamma_0 &=& k^2\frac{\cosh{(L\beta\alpha)}}{2\sinh{(L\alpha)}}
    \left\{1-\beta\frac{\tanh{(L\beta\alpha)}}{\tanh{(L\alpha)}}\right\},
    \\
    \gamma_1 &=& \kperp^2 \left[
      \frac{\cosh{(L\beta\alpha)}-\cosh{(L\alpha)}}{\sinh^3{(L\alpha)}}
      + \frac{\cosh{(L\beta\alpha)}}{2\sinh{(L\alpha)}}
      \left\{1-\beta\frac{\tanh{(L\beta\alpha)}}{\tanh{(L\alpha)}}
      \right\}\right],
    \\
    \gamma_2 &=& (\omega^2-\kpara^2)\left
    [ \frac{1-\beta^2}2\coth{(L\alpha) - \frac{\cosh{(L\beta\alpha)}}
    {2\sinh{(L\alpha)}}
    \left\{1-\beta\frac{\tanh{(L\beta\alpha)}}
     {\tanh{(L\alpha)}}\right\}}\right],
\end{eqnarray}
and
\begin{equation}
    \label{eq:expfac}
    E(\alpha, \beta, k, L) =
    \exp\left[-\alpha+\frac{\alpha(1-\beta^2)}{4} \frac{k^2}{\me^2} +
      \left\{ \frac{\alpha(1-\beta^2)}{4}
        +\frac{\cosh{(L\beta\alpha)}-\cosh{(L\alpha)}}{2L\sinh{(L\alpha)}}
      \right\} \frac{\kperp^2}{\me^2}\right],
\end{equation}
where $e$ denotes electron charge in MKS unit ($e^2 \simeq$ 1/137),
$\me$ is electron mass, and $L (\equiv B/B_c)$ is a dimensionless
magnetic field normalized by the critical magnetic field ($B_c =
\me^2/e$). $f_i^{\mu\nu}$ is expressed as
\begin{eqnarray}
    \label{eq:f_0}
    f_0^{\mu\nu} &=& g^{\mu\nu} - \frac{k^{\mu}k^{\nu}}{k^2}, \\
    \label{eq:f_12}
    f_l^{\mu\nu} &=&
    \frac{b_l^{\mu}b_l^{\nu}}{b_l^{\gamma}b_{l\gamma}},
    \qquad ({\rm for} \ l = 1,2),
\end{eqnarray}
with
\begin{eqnarray}
    \label{eq:b_i}
    b_1^{\mu} &=& F^{\mu\gamma}k_{\gamma},
    \qquad b_2^{\mu} = F^{*\mu\gamma}k_{\gamma}, \\
    b_3^{\mu} &=& k^2 F^{\mu\gamma} F_{\gamma\delta}k^{\delta}
    - k^{\mu}k^{\gamma}F_{\gamma\delta}F^{\delta\epsilon}k_{\epsilon},\\
    b_4^{\mu} &=& k^{\mu},
\end{eqnarray}
where $F^{\mu\nu}$ denotes Maxwell stress 4-tensor and its dual tensor
is defined by $F^{*\mu\nu} \equiv -
\frac12\epsilon^{\mu\nu\gamma\delta}F_{\gamma\delta}$.

It is easy for us to check that only $G_0$ contributes to the vacuum
polarization tensor if the magnetic field is very week $L \ll 1$. In
such a week field limit, we know the form of
$\Pi^{\mu\nu}(k)$ in usual QED as,
\begin{equation}
    \label{eq:B_weak}
    \Pi^{\mu\nu}(k,0) = G_0|_{B=0}\, f_0^{\mu\nu},
\end{equation}
where
\begin{equation}
    G_0|_{B=0} = \frac{e^2}{(2\pi)^2}k^2\int^{\infty}_0 \!\!d\alpha \int^1_0
    \!d\beta \frac{1-\beta^2}{2\alpha}
    \exp\left[-\alpha+\frac{\alpha(1-\beta^2)}{4}\frac{k^2}{\me^2}\right].
    \label{eq:G_0_weak}
\end{equation}
In Eq.~(\ref{eq:G_0_weak}) we should get rid of the divergence at
$\alpha$ = 0 and regularize it. Then we obtain the regularized form of
the vacuum polarization tensor at $L = 0$,
\begin{equation}
    \label{eq:reg_pi}
    {\rm reg}\Pi^{\mu\nu}(k,0) = A(k)f_0^{\mu\nu},
\end{equation}
with
\begin{equation}
    \label{eq:A_k}
    A(k) \equiv {\rm reg}G_0|_{B=0} =  - \frac{e^2}{(2\pi)^2}k^2
     \left[ \frac19 - \frac{(1-h(k))(4\me^2+2k^2)}{3k^2}\right],
\end{equation}
where
\begin{equation}
    \label{eq:h_k}
    h(k) = \left\{
      \begin{array}{ll}
          \sqrt{\frac{4\me^2}{k^2}-1}\cot^{-1}
          \left(\sqrt{\frac{4\me^2}{k^2}-1}\right),
          \qquad ({\rm for} \  0 \le k^2 \lesssim 4\me^2),\\
          \sqrt{1-\frac{4\me^2}{k^2}}\coth^{-1}
          \left(\sqrt{1-\frac{4\me^2}{k^2}}\right),
          \qquad ({\rm for} \  k^2 < 0).
       \end{array}
    \right.
\end{equation}
Thus, to obtain the regularized form of the polarization tensor in
a strong magnetic field ($L \gtrsim 1$), we have only to substitute $G_0$
 with
\begin{equation}
    \label{eq:regg0}
    {\rm reg}G_0 = G_0 - G_0|_{B=0} + A(k).
\end{equation}

\section{refractive indices in a strong magnetic field}
\label{sec:numerical}

As we mentioned in the previous sections, the refractive indices of
photon would deviate from unity in a strong magnetic field
because the vacuum polarization is influenced by the magnetic field
and the dispersion relation is changed. The refractive indices is
defined from the dispersion relation as,
\begin{equation}
    \label{eq:mu2_def}
    \mu^2 = \frac{|\vect{k}|^2}{\omega^2},
\end{equation}
where $\vect{k}$ is a spatial 3-vector of $k$.
To obtain the dispersion relation in a strong magnetic field, we
consider the wave equation of photon,
\begin{equation}
    \label{eq:wave_eq}
    \left[k^2g^{\mu\nu} - k^{\mu}k^{\nu} + 4\pi \ {\rm
      reg}\Pi^{\mu\nu}(k) \right]A_{\mu}(k) = 0.
\end{equation}
In this equation the prefactor of $\Pi^{\mu\nu}(k)$, ``4$\pi$'',
originates in MKS unit. The determinant of the matrix should be zero
so that Eq.~(\ref{eq:wave_eq}) could have a nontrivial solution. When
we choose the radiation gauge $A_{\mu} = (0, \vect{A})$, we get a
quadratic equation and we obtain two solutions,
\begin{equation}
    \label{eq:mu2_1}
    \mu^2_1 = \frac{1+\chi_0}{1+\chi_0-\sin^2\theta \chi_1},
\end{equation}
\begin{equation}
    \label{eq:mu2_2}
    \mu^2_2 = \frac{1+\chi_0+\chi_2}{1+\chi_0 + \cos^2\theta \chi_2},
\end{equation}
where
\begin{eqnarray}
    \label{eq:chi_0}
    \chi_0 &=& \frac{4\pi {\rm reg}G_0}{k^2}, \\
    \label{eq:chi_1}
    \chi_1 &=& \frac{4\pi G_1}{\kperp^2}, \\
    \label{eq:chi_2}
    \chi_2 &=& \frac{4\pi G_2}{\kpara^2}.
\end{eqnarray}
Here $\mu_1^2$ corresponds to the eigen vector $\left(0,1,0\right)$
and $\mu_2^2$ to $\left((1+\chi_0+\chi_2)\cos\theta, \ 0,
  \ - (1+\chi_0)\sin\theta\right)$.

\begin{figure}
   \begin{center}
     \centerline{\psfig{figure=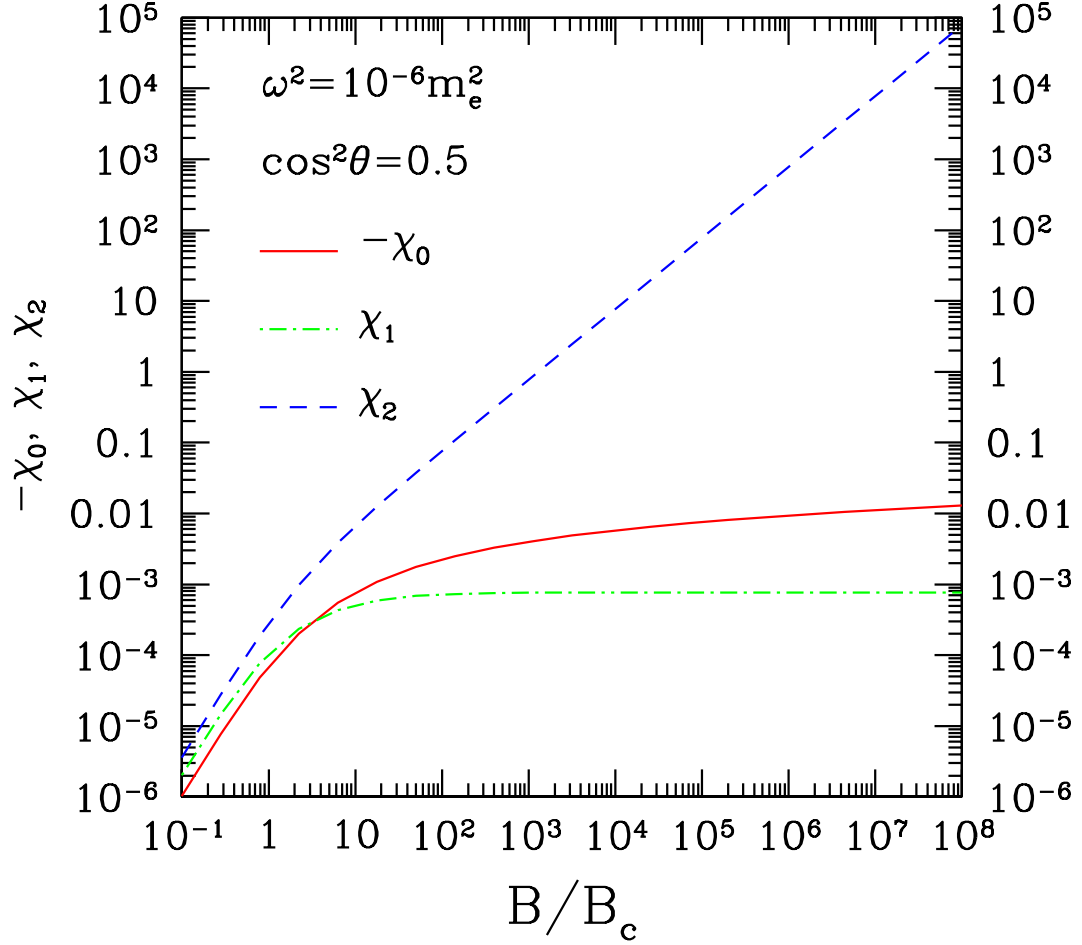,width=17cm}}
       \caption{
       Plots of $-\chi_0$, $\chi_1$ and $\chi_2$ as a function of
       $B/B_c$.  Here we adapt $\omega^2 = 10^{-6} \me^2$ and
       $\cos^2\theta = 0.5$. We find that the magnitude of $\chi_0$
       increases as $B/B_c$ increases ($\propto \log(B/B_c)$),
       $\chi_1$ reaches the limit value ($\sim$ 7.7 $\times 10^{-4}$),
       and $\chi_2 \simeq \frac{e^2}{3\pi}B/B_c$ for $B/B_c \gg 1$.  }
       \label{fig:chi012}
   \end{center}
\end{figure}

Since each $\chi_i$ depends on $\mu^2$ through $k$ and $G_i$,
Eqs.~(\ref{eq:mu2_1}) and (\ref{eq:mu2_2}) are implicit equations for
$\mu^2$. In addition, in the literatures,
e.g.~\cite{melrose_stoneham}, they gave only the integral form and
some limit cases. Hence we must solve the equations numerically to
obtain values in the entire parameter space.  In
Fig.~\ref{fig:chi012}, we plot $- \chi_0$ as a function of $L$
(=$B/B_c$) in a low energy limit ($\omega^2 = 10^{-6} \me^2$). From
the plot, we find that the magnitude of $\chi_0$ increases as $L$
increases. Thus the contribution from $\chi_0$ to the refractive
indices in Eqs.~(\ref{eq:mu2_1}) would not be negligible in an
extremely strong magnetic field. This behavior agrees with our
analytical estimation that $\chi_0 \propto - \log(L)$ in a strong
magnetic field limit ($L \to \infty$) and contradicts with statement
by Melrose and Stoneham~\cite{melrose_stoneham} that $\chi_0 \propto
\exp[-L]$ for this limit. This difference, however, is substantial only
for extremely large magnetic fields. Nonetheless, we did not drop this
term in estimating the correct refractive indices. In
Fig.~\ref{fig:chi012} we also plot $\chi_1$ and $\chi_2$ as a function
of $L$. $\chi_1$ approaches the limit value ($\sim 7.7 \times
10^{-4}$) for $L \gg 1$. This feature is consistent with our
analytical estimation and again disagrees with Melrose and
Stoneham~\cite{melrose_stoneham} who claimed that $\chi_1 \propto
\exp[-L]$ for this limit. As for $\chi_2$, it is found that $\chi_2$
is linearly proportional to $L$ in the strong magnetic field and the
weak energy limit. It agrees exactly with our analytical estimation
that $\chi_2 \simeq \left(e^2/3\pi\right) L$ under the condition that
$L \gg 1$, $\omega^2 - \kpara^2 \ll 4\me^2$, and $\kperp^2 \ll
2\me^2L$~\cite{melrose_stoneham}.

\begin{figure}
   \begin{center}
     \centerline{\psfig{figure=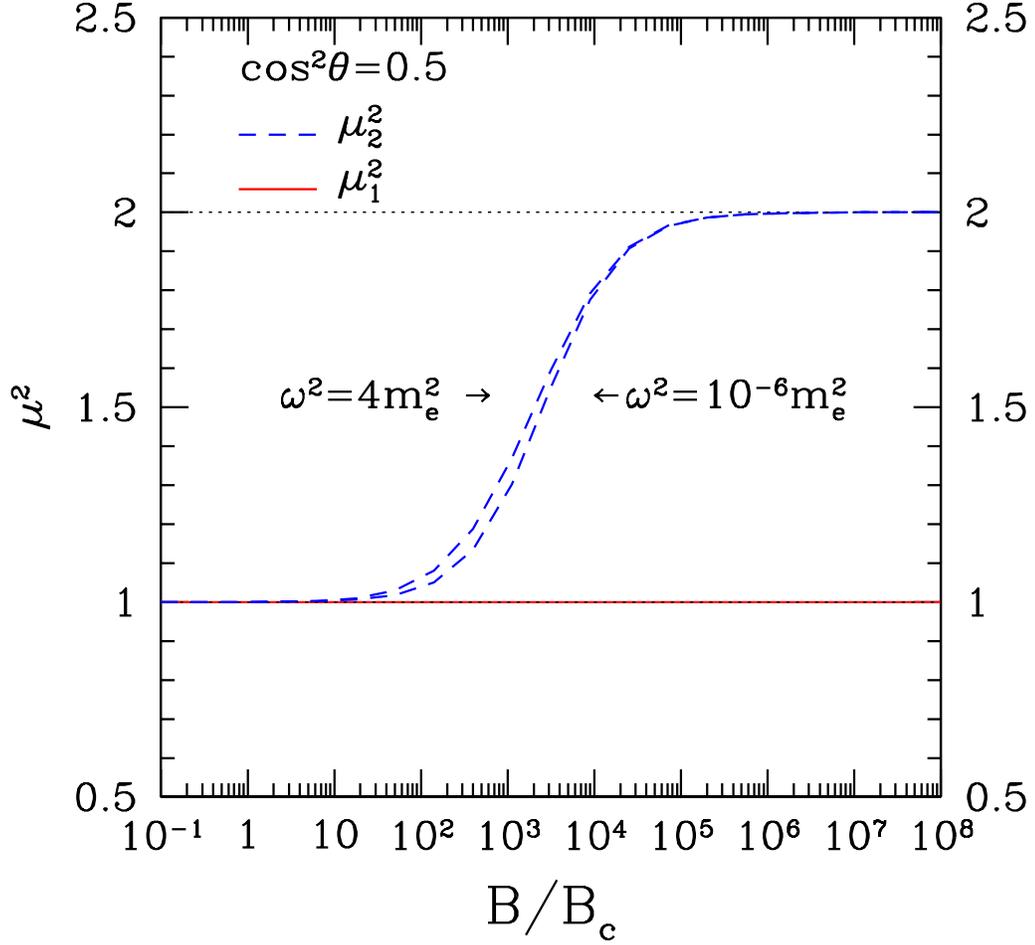,width=17cm}}
       \caption{
       Plot of the refractive indices as a function of $B/B_c$.  The
       solid line represents $\mu^2_1$ and the dashed line represents
       $\mu^2_2$. The left (right) dashed line is the case for
       $\omega^2 = 4\me^2 \ (10^{-6} \me^2)$. Here we adapt $\omega^2
       = 10^{-6} \me^2$ and $\cos^2\theta = 0.5$. We find that
       $\mu^2_2$ reaches the limit value ($1/\cos^2\theta = 2$) in the
       strong $B$ limit.}
       \label{fig:L_mu}
   \end{center}
\end{figure}

\begin{figure}
   \begin{center}
     \centerline{\psfig{figure=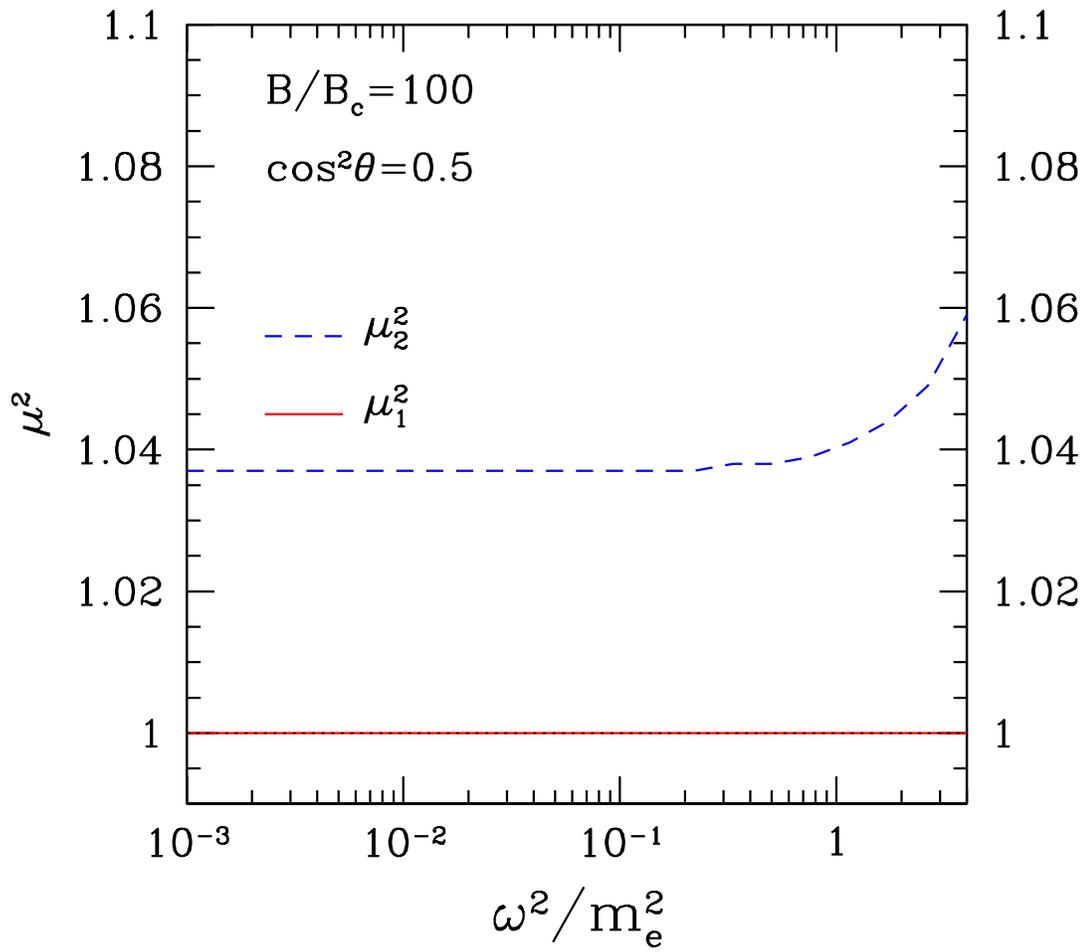,width=17cm}}
       \caption{
       Plot of the refractive indices as a function of
       $\omega^2/\me^2$. The solid line represents $\mu^2_1$ and the
       dashed line represents $\mu^2_2$. Here we adapt $B/B_c = 10^2$
       and $\cos^2\theta = 0.5$.}
       \label{fig:w_mu}
   \end{center}
\end{figure}

\begin{figure}
   \begin{center}
     \centerline{\psfig{figure=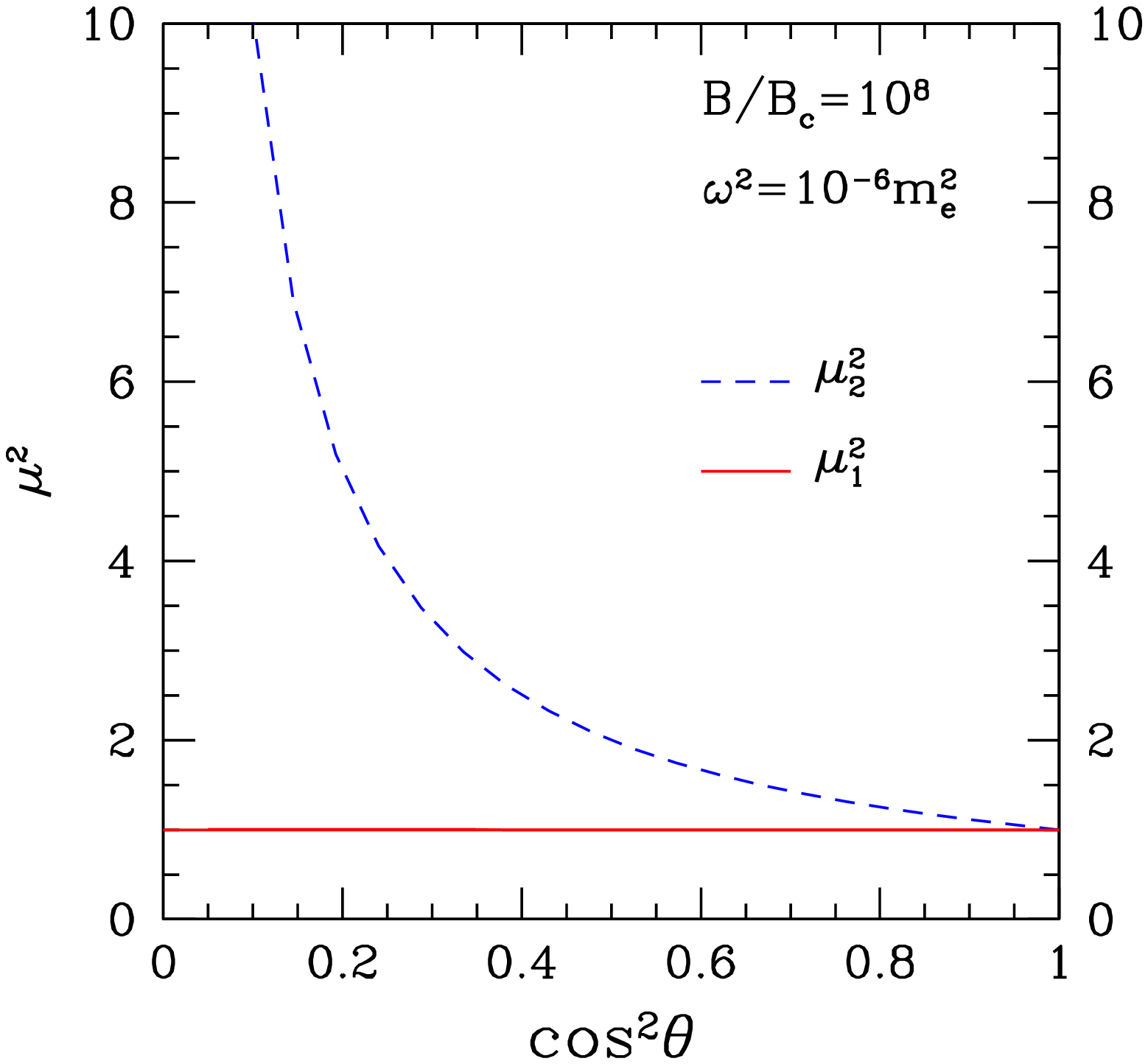,width=17cm}}
       \caption{
       Plot of the refractive indices as a function of $\cos^2\theta$.
       The solid line represents $\mu^2_1$ and the dashed line
       represents $\mu^2_2$. Here we adapt $B/B_c = 10^8$ and
       $\omega^2 = 10^{-6} \me^2$. We find that $\mu_2^2$ scales
       $1/\cos^2\theta$ in such a strong $B$ limit.}
       \label{fig:c2_mu}
   \end{center}
\end{figure}

In Fig.~\ref{fig:L_mu} we plot the obtained refractive indices as a
function of $L$.  The solid and dashed lines represent $\mu^2_1$ and 
$\mu^2_2$, respectively. It is easy for us to understand the
behavior of $\mu^2$s in a strong magnetic field. From
Eq.~(\ref{eq:mu2_1}) and~(\ref{eq:mu2_2}), we find that
\begin{eqnarray}
    \label{eq:mu_L_limit_1}
    \mu^2_1 &\simeq& 1 + 7.7 \times 10^{-4} \sin ^{2} \theta, \\
    \label{eq:mu_L_limit_2}
    \mu^2_2 &\simeq& 1/\cos^2\theta,
\end{eqnarray}
for $L \gg 1$ in the case of the weak energy and as long as $\chi_0$ is 
much smaller than unity. The photon-energy
dependence of $\mu^2$ is shown in Fig.~\ref{fig:w_mu}. In this plot we
find that only near the threshold ($\omega^2 \sim 4\me^2$) the energy
dependence of $\mu^2_2$ becomes important. $\mu^2_1$ is insensitive to the
photon energy. In Fig.~\ref{fig:c2_mu} we plot $\mu^2$ as a function
of $\cos^2\theta$ in a strong magnetic field ($L = 10^8$). It is clear
that $\mu_2^2$ is proportional to 1/$\cos^2\theta$.

Here we give the fitting formula of $\mu_2^2$,
\begin{equation}
    \label{eq:fitting_mu2_2}
    \mu_2^2 = C_1\left(\tanh\left[C_2(x - C_3)\right] + 1\right)+ 1,
\end{equation}
with
\begin{equation}
    \label{eq:x_logL}
    x = \log_{10}(B/B_c),
\end{equation}
and
\begin{equation}
    \label{eq:coef}
    \left\{
      \begin{array}{ll}
          C_1 = \tan^2\theta/2, \\
          C_2 = 1.15 -
          7.07\times10^{-3}\left(\omega^2/\me^2\right)^{1.60}
          /\sqrt{\cos^2\theta}, \\
          C_3 = 3.11 -\log_{10}\left(\cos^2\theta \right) -
          1.84\times10^{-2}\left(\omega^2/\me^2\right)/
          \sqrt{\cos^2\theta}.
      \end{array}
    \right.
\end{equation}
This fitting formula reproduces the numerical results within the error 
of less than 10 $\%$ for a wide parameter range 
(0.1 $\le \cos^2\theta \le$ 1, $\omega^2/\me^2
\lesssim$ 4, and $0 \le B/B_c \lesssim 10^{10}$). In the
large $L$ limit especially, Eq.~(\ref{eq:fitting_mu2_2}) approaches 
the value of the analytical estimation ($\simeq 1/\cos^2\theta$). 
In addition, the low energy limit of the fitting formula 
($\omega^2 \to 0$) agrees with the
numerical estimations very well within less than 1 $\%$.

\section{retarded polarization tensor in finite temperature plasmas}
\label{sec:plasm}

In the previous sections, we have considered the vacuum polarization 
tensor. It is also our concern to calculate the polarization tensor 
for plasmas with finite temperatures. We will extend the previous 
formulation to the finite density and temperature case. We will rely 
on the real time formalism of the finite density and temperature 
field theory and obtain the expression for the retarded polarization
tensor. Recently, some authors~\cite{gk} gave the expression of the
chronological polarization tensor on a similar footing. Although two 
polarizations are related to each other, the retarded one allows more 
direct physical interpretation.Moreover, the retarded polarization 
tensor should be obtained by analytical extension of the
imaginary time polarization tensor which has been given by the
author~\cite{al}.

It is known from the finite density and temperature field theory that the
retarded polarization tensor is expressed as
\begin{equation}
i \Pi_{r}^{\mu\nu}(x, x') = e^{2}\,{\rm Tr} \left [\gamma ^{\mu} G_{c}(x, x')
\gamma^{\nu}
G_{r}(x', x) + \gamma ^{\mu} G_{a}(x, x') \gamma^{\nu} G_{c}(x', x) \right ]
\quad .
\label{eq:pir}
\end{equation}
In the above equation, the subscripts $r$, $a$, $c$ denote the retarded,
advanced and Keldysh components of Green function, 
respectively.~\cite{ch} For the stationary system, we can in general 
assume that the vector potential is time independent. In
this case the polarization tensor $\Pi_{r}(x, x')$ is a function of the time
difference $t - t'$ alone. It is possible that it depends on the spatial
coordinates
{\boldmath $x$} and {\boldmath $x'$} separately. Fourier transforming the
polarization tensor with respect to $t - t'$, we obtain
\begin{equation}
i\tilde{\Pi}_{r}(p_{0}) = e^{2}\int \frac{dq_{0}}{2\pi} \,{\rm Tr} \left [
\gamma ^{\mu} \tilde{G}_{c}(q_{0} + p_{0}) \gamma ^{\nu}
\tilde{G}_{r}(q_{0})
+ \gamma ^{\mu} \tilde{G}_{a}(q_{0} + p_{0}) \gamma ^{\nu}
\tilde{G}_{c}(q_{0})
\right ] \quad .
\label{eq:fourier}
\end{equation}
In the above equation, the tilde means the Fourier component and the spatial
coordinates are dropped for simplicity. Noting that the retarded and
advanced Green functions for the finite density and temperature are 
identical to the counterparts for vacuum, we obtain
\begin{equation}
\tilde{G}_{r,a}(p_{0}) = \frac{1}{2\pi i} \int d \omega
\frac{\tilde{G}^{vac}_{F}(\omega) - \tilde{G}^{vac}_{\tilde{F}}(\omega)}
{\omega - p_{0} \mp i \varepsilon} \quad ,
\label{eq:pirel}
\end{equation}
where the upper and lower signs correspond to the retarded and advanced
Green functions, respectively. The subscripts $F$ and $\tilde{F}$ stand for
the chronological and antichronological Green functions, respectively, for
magnetized vacuum, which are calculated by the Schwinger's proper time
method as shown above. It is also known that the Keldysh component of Green
function can be obtained from the retarded and advanced Green functions and
the distribution function by the following relation:
\begin{equation}
\tilde{G}_{c}(p_{0}) = \tilde{G}_{r}(p_{0}) [1 - 2f(p_{0})] -
[1 - 2 f(p_{0})] \tilde{G}_{a}(p_{0}) \quad .
\label{eq:gkeldysh}
\end{equation}
Combining this with Eq.~(\ref{eq:pirel}), we can obtain the Keldysh
component
of Green function also from the chronological and anti-chronological Green
functions
for vacuum.

Putting Eqs.~(\ref{eq:pirel}), (\ref{eq:gkeldysh}) into
Eq.~(\ref{eq:fourier})
and using the relations
\begin{eqnarray}
\tilde{G}_{r}(q_{0}) & = & \Theta (q_{0}) \tilde{G}_{F}(q_{0}) -
\Theta(-q_{0}) \tilde{G}_{\tilde{F}}(q_{0}) \\
\tilde{G}_{a}(q_{0}) & = & \Theta (q_{0}) \tilde{G}_{\tilde{F}}(q_{0}) -
\Theta(-q_{0}) \tilde{G}_{F}(q_{0}) \quad ,
\end{eqnarray}
we finally obtain
\begin{eqnarray}
i \tilde{\Pi}_{r}(p_{0}) =  \int \frac{dq_{0}}{2\pi} [1 - 2f(q_{0} + p_{0})]
 \times & {\rm Tr} &
\left[ \Theta (q_{0}) \gamma ^{\mu}\tilde{G}_{F}(q_{0} + p_{0})
\gamma ^{\nu} \tilde{G}_{F}(q_{0})
\right . \nonumber \\& - & \ 
\Theta (q_{0}) \gamma ^{\mu} \tilde{G}_{\tilde{F}}(q_{0} + p_{0})
\gamma ^{\nu}\tilde{G}_{F}(q_{0}) 
\nonumber \\
& - & \ \Theta(-q_{0}) \gamma ^{\mu}\tilde{G}_{F}( q_{0} +
p_{0})\gamma ^{\nu}\tilde{G}_{\tilde{F}}(q_{0})
\nonumber \\ & + & \, \left .
\Theta (-q_{0}) \gamma ^{\mu} \tilde{G}_{\tilde{F}}( q_{0} + p_{0}) 
\gamma ^{\nu} \tilde{G}_{\tilde{F}} (q_{0}) \right ] \nonumber \\
 - \int \frac{dq_{0}}{2\pi} [1 - 2f(q_{0} - p_{0})] \times &{\rm Tr}&
\left[\Theta (-q_{0}) \gamma ^{\mu} \tilde{G}_{F}(q_{0}) 
\gamma ^{\nu} \tilde{G}_{F}(q_{0} - p_{0}) \right .
\nonumber \\
& - & \ \Theta (-q_{0}) \gamma ^{\mu} \tilde{G}_{F}(q_{0})
\gamma ^{\nu} \tilde{G}_{\tilde{F}}(q_{0} - p_{0}) \nonumber \\
& - & \ \Theta (q_{0}) \gamma ^{\mu} \tilde{G}_{\tilde{F}}(q_{0})
\gamma ^{\nu} \tilde{G}_{F}(q_{0} - p_{0}) \nonumber \\
& + & \ \left . \Theta (q_{0}) \gamma ^{\mu} \tilde{G}_{\tilde{F}}(q_{0})
\gamma ^{\nu} \tilde{G}_{\tilde{F}}(q_{0} - p_{0}) \right ] \quad .
\label{eq:pifin}
\end{eqnarray}
One easily recognizes that the structure of the integrand is quite similar
to the vacuum  polarization tensor apart from the integration over $q_{0}$ and
the various  combinations of $\tilde{G}_{F}$ and $\tilde{G}_{\tilde{F}}$.
This enables us to simplify the integrand along the same line as for the
vacuum case.

Denoting as $^{FF}\Pi_{r}$ the contribution from the terms containing the
product $G_{F}G_{F}$ and similarly for the other contributions, we calculate
separately those terms. The antichronological Green function 
$G_{\tilde{F}}$ is obtained by taking the integration region 
$[0, -\infty]$ instead of $[0, \infty]$ in the Schwinger's proper 
time formalism :
\begin{equation}
G_{\tilde{F}}(x, x') = \frac{1}{i} \int^{-\infty}_{0}ds \exp
\left[-i \left( m^{2} - (\gamma^{\mu} \Pi_{\mu})^{2} \right )s \right ]
(\gamma^{\mu}\Pi_{\mu} + m) \quad .
\label{eq:anti}
\end{equation}
Following Stoneham, we can simplify this equation. Assuming that the vector
potential is time independent and $A_{0} = 0$, we can Fourier 
transform $G_{F}$ and $G_{\tilde{F}}$ with respect to $t-t'$. 
Plugging them into the definition of the polarization tensor, one sees 
that the gauge dependent terms cancel out just like the vacuum case and 
the polarization tensor becomes the function of the difference
of the spatial coordinates, which then makes it possible for us to take the
Fourier transformation with respect to the spatial coordinates. We 
finally obtain for the contribution from the terms with the product 
$G_{F}G_{F}$
\begin{eqnarray}
^{FF}\Pi_{r}^{\mu\nu}(p) & = & \int_{-\infty}^{\infty}
\frac{dk_{0}}{4\pi m} FF(k_{0}, p_{0})
\frac{e^{2}m^{2}}{(2\pi)^{2}}(1 - i) \sqrt{\frac{2\pi}{L}} \nonumber \\
& \times & \int^{\infty}_{0} \frac{d\alpha}{\sqrt{\alpha}}
\int^{\alpha}_{-\alpha} d\beta
\exp \left [-i\frac{\alpha}{L}\right] \exp 
\left[ - i \frac{\alpha ^{2} -
\beta ^{2}}{4\alpha L} \frac{k_{z}^{2}}{m^{2}} \right ] \nonumber \\
& \times & \exp \left [-i \frac{\cos \beta - \cos \alpha}
{2 L \sin \alpha}
\frac{k_{\perp}^{2}}{m^{2}} \right ] \exp \left [i \left\{
\frac{\alpha + \beta}{2L}
\frac{(k_{0} + p_{0})^{2}}{m^{2}} + \frac{\alpha - \beta}{2 L}
\frac{k_{0}^{2}}{m^{2}} \right \} \right ] \nonumber \\
& \times & d^{\mu\nu} \quad .
\label{eq:finpi}
\end{eqnarray}
Here $FF(k_{0}, p_{0})$ is an abbreviation for
the following function,
\begin{eqnarray}
FF(k_{0}, p_{0}) & = & \Theta (k_{0}) \left [1 - 2 f(k_{0} + p_{0}) \right ]
- \Theta (-k_{0} - p_{0}) \left [1 - 2 f(k_{0}) \right ] \quad ,\\
& = & \Theta (k_{0})\ \ \ \Theta (k_{0} + p_{0}) \ \
\left [1 - 2f_{e}(k_{0} + p_{0}) \right ]
\nonumber \\
& + & \Theta (-k_{0}) \, \Theta (-k_{0} - p_{0})
\left [1 - 2f_{e^{+}}(|k_{0}|) \right ]
\nonumber \\
& - & \Theta (k_{0}) \ \ \ \Theta (-k_{0} - p_{0})
\left [1 - 2f_{e^{+}}(|k_{0} + p_{0}|)
+ 1 - 2 f_{e}(k_{0}) \right ] \quad ,
\label{eq:defff}
\end{eqnarray}
where $f_{e}$ and $f_{e^{+}}$ are Fermi-Dirac distribution functions for
electron
and positron, respectively.
Except for the integral over the distribution functions, the resemblance of
the
Eq.~(\ref{eq:finpi}) to the vacuum counter part is clear. The remaining
factor
$d^{\mu \nu}$, which is symmetric with respective to the superscripts, is
given as
\begin{eqnarray}
d^{01} & = & -\frac{1}{2} \frac{\cos \beta - \cos \alpha}{\sin \alpha}
\left \{ \left [ \cot \left (\frac{\alpha - \beta}{2} \right ) - \cot \alpha
\right ]
\frac{k_{0} + p_{0}}{m} \left [ \cot \left (\frac{\alpha + \beta}{2}
\right )
\frac{k_{x}}{m} + \frac{k_{y}}{m} \right ]\right . \nonumber \\
&& \qquad \qquad \qquad \quad +  \left . \left [
\cot \left (\frac{\alpha + \beta}{2} \right ) - \cot \alpha \right ]
\frac{k_{0}}{m} \left [\cot \left (
\frac{\alpha - \beta}{2} \right ) \frac{k_{x}}{m} - \frac{k_{y}}{m} \right ]
\right \} \quad , \\
d^{02} & = & -\frac{1}{2} \frac{\cos \beta - \cos \alpha}{\sin \alpha}
\left \{ \left [ \cot \left (\frac{\alpha - \beta}{2} \right ) - \cot \alpha
\right ]
\frac{k_{0} + p_{0}}{m} \left [ \cot \left (\frac{\alpha + \beta}{2}
\right )
\frac{k_{y}}{m} - \frac{k_{x}}{m} \right ]\right . \nonumber \\
&& \qquad \qquad \qquad \quad +  \left . \left [
\cot \left (\frac{\alpha + \beta}{2} \right ) - \cot \alpha \right ]
\frac{k_{0}}{m} \left [\cot \left (
\frac{\alpha - \beta}{2} \right ) \frac{k_{y}}{m} + \frac{k_{x}}{m} \right ]
\right \} \quad , \\
d^{03} & = & - \frac{1}{2} \cot \alpha \frac{k_{z}}{m} \left [
\frac{2k_{0}}{m} + \frac{p_{0}}{m} \left (1 + \frac{\beta}{\alpha} \right )
\right ] \quad , \\
d^{12} & = & \frac{\sin \beta}{\sin \alpha} \left [\frac{k_{0} + p_{0}}{m}
\frac{k_{0}}{m} + 1 - \frac{i}{2} \frac{L}{\alpha} -
\frac{\alpha ^{2} - \beta ^{2}}{4 \alpha ^{2}} \frac{k_{z} ^{2}}{m^{2}}
\right ]
+ \frac{\cos \alpha - \cos \beta }{\sin ^{3} \alpha }
\frac{k_{x}}{m} \frac{k_{y}}{m} \quad , \\
d^{13} & = & - \frac{1}{2} \frac{\cos \beta}{\sin \alpha} \left [
1 - \frac{\beta}{\alpha}\frac{\tan \beta}{\tan \alpha} \right ]
\frac{k_{x}}{m} \frac{k_{z}}{m} \quad , \\
d^{23} & = & - \frac{1}{2} \frac{\cos \beta}{\sin \alpha} \left [
1 - \frac{\beta}{\alpha}\frac{\tan \beta}{\tan \alpha} \right ]
\frac{k_{y}}{m} \frac{k_{z}}{m} \quad , \\
d^{00} & = & - \cot \alpha \left [\frac{k_{0} + p_{0}}{m}\frac{k_{0}}{m} - 1
+
\frac{L}{2 \alpha} i +  \frac{\alpha ^{2} - \beta ^{2}}{4 \alpha
^{2}}
\frac{k_{z}^{2}}{m^{2}} \right ] \nonumber \\
&& \qquad \qquad \qquad \qquad \qquad + \frac{L}{\sin ^{2} \alpha} i
- \frac{1}{2} \frac{\cos \beta - \cos \alpha}{\sin ^{3} \alpha }
\left ( \frac{k_{x}^{2}}{m^{2}} + \frac{k_{y}^{2}}{m^{2}} \right )
\quad , \\
d^{11} & = & - \frac{\cos \beta}{\sin \alpha} \left [
\frac{k_{0} + p_{0}}{m} \frac{k_{0}}{m} + 1 - \frac{L}{2 \alpha} i
- \frac{\alpha ^{2} - \beta ^{2}}{4 \alpha ^{2}} \frac{k_{z}^{2}}{m^{2}}
\right ]
\nonumber \\ && \qquad \qquad \qquad \qquad \qquad - \frac{\cos \beta - \cos
\alpha }
{2 \sin ^{3} \alpha} \left [ \frac{k_{x}^{2}}{m^{2}}
- \frac{k_{y}^{2}}{m^{2}} \right ] \quad , \\
d^{22} & = & - \frac{\cos \beta}{\sin \alpha} \left [
\frac{k_{0} + p_{0}}{m} \frac{k_{0}}{m} + 1 - \frac{L}{2 \alpha} i
- \frac{\alpha ^{2} - \beta ^{2}}{4 \alpha ^{2}} \frac{k_{z}^{2}}{m^{2}}
\right ]
\nonumber \\ && \qquad \qquad \qquad \qquad \qquad + \frac{\cos \beta - \cos
\alpha }
{2 \sin ^{3} \alpha} \left [ \frac{k_{x}^{2}}{m^{2}}
- \frac{k_{y}^{2}}{m^{2}} \right ] \quad , \\
d^{33} & = & - \cot \alpha \left [\frac{k_{0} + p_{0}}{m}\frac{k_{0}}{m} + 1
+
\frac{L}{2 \alpha} i  + \frac{\alpha ^{2} - \beta ^{2}}{4 \alpha
^{2}}
\frac{k_{z}^{2}}{m^{2}} \right ] \nonumber \\
&& \qquad \qquad \qquad \qquad \qquad - \frac{L}{\sin ^{2} \alpha} i
+ \frac{1}{2} \frac{\cos \beta - \cos \alpha}{\sin ^{3} \alpha }
\left ( \frac{k_{x}^{2}}{m^{2}} + \frac{k_{y}^{2}}{m^{2}} \right )
\quad .
\label{eq:eqdmn}
\end{eqnarray}
In the above equation, the integral region for $\beta$ can be changed from
$[-\alpha , \alpha]$ to $[0, \alpha]$ under the recognition that we take
only the even part of the integrand with respect to $\beta$.

The other contributions to $\Pi_{r}$ with different combinations of $G_{F}$
and
$G_{\tilde{F}}$ are obtained in the same way. It turns out that the
resultant
equations
are obtained from Eq.~(\ref{eq:finpi}) with a change of the integral region
and
a replacement of the distribution functions as shown below:
\begin{eqnarray}
^{F\tilde{F}}\Pi_{r}^{\mu\nu} &:& FF(k_{0}, p_{0}) \rightarrow
F\tilde{F}(k_{0}, p_{0}) , \quad \mbox{integral region}
\rightarrow \int_{-\infty}^{\infty}d\alpha \int^{\infty}_{\alpha}d\beta
\quad , \\
^{\tilde{F}F}\Pi_{r}^{\mu\nu} &:& FF(k_{0}, p_{0}) \rightarrow
\tilde{F}F(k_{0}, p_{0}) , \quad \mbox{integral region}
\rightarrow \int_{-\infty}^{\infty}d\alpha \int_{-\infty}^{-\alpha}d\beta
\quad , \\
^{\tilde{F}\tilde{F}}\Pi_{r}^{\mu\nu} &:& FF(k_{0}, p_{0}) \rightarrow
\tilde{F}\tilde{F}(k_{0}, p_{0}) , \quad \mbox{integral region}
\rightarrow \int_{-\infty}^{0}d\alpha \int^{\alpha}_{-\alpha}d\beta
\quad .
\label{eq:rep}
\end{eqnarray}
In the above equations, the phase should be taken as $\sqrt{-\alpha} = -i
\sqrt{|\alpha|}$, and the factors involving distribution functions are given
as
\begin{eqnarray}
F\tilde{F}(k_{0}, p_{0}) & = & -\Theta (-k_{0}) \left [1 - 2 f(k_{0} +
p_{0}) \right ]
- \Theta (-k_{0} - p_{0}) \left [1 - 2 f(k_{0}) \right ] \quad , \\
\tilde{F}F(k_{0}, p_{0}) & = & -\Theta (k_{0}) \ \ \,
\left [1 - 2 f(k_{0} + p_{0}) \right ]
- \Theta (k_{0} + p_{0}) \ \ \left [1 - 2 f(k_{0}) \right ] \quad , \\
\tilde{F}\tilde{F}(k_{0}, p_{0}) & = & \ \ \, \Theta (-k_{0})
\left [1 - 2 f(k_{0} + p_{0}) \right ]
+ \Theta (-k_{0} - p_{0}) \left [1 - 2 f(k_{0}) \right ] \quad .
\label{eq:defffs}
\end{eqnarray}
Thus the polarization tensor is give by the sum of these terms:
\begin{equation}
\Pi_{r}^{\mu\nu} \ = \ ^{FF}\Pi_{r}^{\mu\nu}
\ + \ ^{F\tilde{F}}\Pi_{r}^{\mu\nu} \ + \ ^{\tilde{F}F}\Pi_{r}^{\mu\nu}
\ + \ ^{\tilde{F}\tilde{F}}\Pi_{r}^{\mu\nu}  \quad .
\label{eq:pisum}
\end{equation}

Although the final form is similar to the vacuum counter part, the numerical
evaluation is very difficult. This is so for the same reason as for the
numerical
evaluation of the vacuum polarization tensor above the threshold of the pair
creation.
This prevents us from performing a Wick rotation for the $\alpha $ and
$\beta $ integrations in Eq.~(\ref{eq:finpi}). These should be the next
step.

\section{Conclusion}
\label{sec:conclusion}
We have numerically calculated the vacuum polarization tensor for various
strengths of the background magnetic field and photon energies below the
threshold of the pair creation. We also varied the propagation directions of
photon. We have obtained the fitting formula which reproduces the numerical
results within 10\% of error. We have also presented the expression
of the retarded polarization tensor for the finite density and temperature,
simplifying some integrations.

Our final goal is to calculate not only the polarization tensor but also
other physical quantities under strong magnetic fields such as the photon
splitting rates and equation of states for a wide range of parameters. 
This paper is the first step for this project.

\acknowledgments

This work is partially supported by the 
Grants-in-Aid by the Ministry of Education, Science, Sports and 
Culture of Japan (No.12740138).


\newpage
%
%
%
\newpage
\newpage
\newpage
\newpage


\begin{references}

\bibitem{du} R.C. Duncan, Proceedings of Fifth Huntsville Gamma-Ray
 Burst Symposium, astro-ph/0002442.

\bibitem{lai} D. Lai, accepted for publication in
 Rev. Mod. Phys. (2001), astro-ph/0009333. 

\bibitem{ko} C. Kouveliotou et al., Nature {\bf 393}, 235 (1998).

\bibitem{dt} R.C. Duncan and C. Thompson, Astrophys. J. Lett. {\bf 392},
 L9 (1992).

\bibitem{pa} B. Paczy\'{n}ski,  in ``The Largest Explosions Since the Big
     Bang: Supernovae and Gamma Ray Bursts", 
     eds M. Livio, K. Sahu, and N. Panagia (Cambridge
     University Press, Cambridge), astro-ph/9909048.

\bibitem{bz} R.D. Blandford, R.L. Znajek, Mon. Not. R. Astron. Soc. 
    {\bf 179}, 433 (1977).

\bibitem{wh} J.C. Wheeler, in ``The Largest Explosions Since the Big
     Bang: Supernovae and Gamma Ray Bursts'', 
     eds M. Livio, K. Sahu, and N. Panagia (Cambridge
     University Press, Cambridge), astro-ph/9909096.

\bibitem{sy} E.M.D. Symbalisty, Astrophys. J. {\bf 285}, 729 (1984).

\bibitem{ks} A. Kusenko and G. Segre, Phys. Rev. D {\bf 59}, 061302 (1999).

\bibitem{hl} C.J. Horowitz and G. Li, Phys. Rev. D {\bf 61}, 063002 (2000).

\bibitem{ad} S.L. Adler, Ann. Phys. {\bf 67}, 599 (1971).

\bibitem{melrose_stoneham}
     D.B. Melrose and R.J. Stoneham, Nuovo Cimento {\bf 32}, 435 (1976).

\bibitem{mz} P. M\'{e}sz\'{a}ros, ``High-Energy Radiation from
     Magnetized Neutron Satrs'', (University of Chicago Press, Chicago).

\bibitem{td} C. Thompson and R.D. Duncan, Mon. Not. R. Astron. Soc. 
    {\bf 275}, 255 (1995).

\bibitem{bm} T. Bulik and M.C. Miller, Mon. Not. R. Astron. Soc. {\bf 288},
    596 (1997).

\bibitem{schwinger} J. Schwinger, \pr {\bf 82}, 664 (1951).

\bibitem{gk} A. Ganguly and S. Konar, Phys. Rev. D {\bf 63}, 065001 (2001).

\bibitem{al} J. Alexander, hep-th/0009204 

\bibitem{ch} K.-C. Chou, Z.-B. Su, B.-L. Hao and L. Yu, Phys. Rep. 
    {\bf 118}, 1 (1985). 

\end{references}
\end{document}